\documentclass[11pt]{article}

\usepackage{fullpage,wrapfig}
\usepackage{amsmath, amsthm, amssymb, algorithm,thmtools,algpseudocode,enumerate,caption,framed}
\usepackage{paralist, sidecap}
\usepackage[usenames,dvipsnames,svgnames,table]{xcolor}
\definecolor{darkgreen}{rgb}{0.0,0,0.9}
\usepackage[colorlinks=true,pdfpagemode=UseNone,citecolor=Blue,linkcolor=Red,urlcolor=Black,pagebackref]{hyperref}
\usepackage{tikz}
\usepackage[T1]{fontenc}
\usepackage[utf8]{inputenc}
\usepackage{authblk}

\newtheorem{theorem}{Theorem}[section]

\newtheorem{lemma}{Lemma}[section]

\newcommand{\ptas}{$\mathsf{PTAS}$ }

\newcommand{\both}{\mathsf{Both}}
\newcommand{\one}{\mathsf{One}}

\newcommand{\mfc}{$\mathsf{LSC}$ }
\newcommand{\fpt}{$\mathsf{FPT}$ }

\newcommand{\apx}{$\mathsf{APX}$}
\newcommand{\elf}{$\mathsf{L}$}
\newcommand{\np}{$\mathsf{NP}$}

\newcommand{\mvc}{$\mathsf{MVC}$ }

\title{Approximability of Covering Cells with Line Segments\thanks{Paz Carmi is supported by Grant 2016116 from the United States-Israel Binational Science Foundation. Anil Maheshwari is supported in part by Natural Sciences and Engineering Research Council of Canada (NSERC). Saeed Mehrabi is supported by a Carleton-Fields postdoctoral fellowship.}}

\author[1]{Paz Carmi}
\author[2]{Anil Maheshwari}
\author[2]{Saeed Mehrabi}
\author[3]{Lu\'{i}s Fernando Schultz Xavier da Silveira}
\affil[1]{{\small Department of Computer Science,

Ben-Gurion University of the Negev, Beer-Sheva, Israel.

			\texttt{carmip@cs.bgu.ac.il}}}
\affil[2]{{\small School of Computer Science,

Carleton University, Ottawa, Canada.

			\texttt{anil@scs.carleton.ca, saeed.mehrabi@carleton.ca}}}
\affil[3]{{\small School of Computer Science and Electrical Engineering,

University of Ottawa, Ottawa, Canada.

\texttt{schultz@ime.usp.br}}}

\date{}

\begin{document}

\maketitle

\begin{abstract}
In COCOA 2015, Korman et al. studied the following geometric covering problem: given a set $S$ of $n$ line segments in the plane, find a minimum number of line segments such that every cell in the arrangement of the line segments is covered. Here, a line segment $s$ \emph{covers} a cell $f$ if $s$ is incident to $f$. The problem was shown to be \np-hard, even if the line segments in $S$ are axis-parallel, and it remains \np-hard when the goal is cover the ``rectangular'' cells (i.e., cells that are defined by exactly four axis-parallel line segments).

In this paper, we consider the approximability of the problem. We first give a \ptas for the problem when the line segments in $S$ are in any orientation, but we can only select the covering line segments from one orientation. Then, we show that when the goal is to cover the rectangular cells using line segments from both horizontal and vertical line segments, then the problem is \apx-hard. We also consider the parameterized complexity of the problem and prove that the problem is \fpt when parameterized by the size of an optimal solution. Our \fpt algorithm works when the line segments in $S$ have two orientations and the goal is to cover \emph{all} cells, complementing that of Korman et al.~\cite{KormanPR18} in which the goal is to cover the ``rectangular'' cells.
\end{abstract}

%%%%%%%%%%%% NEW SECTION %%%%%%%%%%%%%%%%%
\section{Introduction}
\label{sec:introduction}
\emph{Set Cover} is a well-studied problem in computer science. The input to the problem is a ground set $\mathcal{G}$ of $n$ elements and a set $\mathcal{S}$ of $m$ subsets of $\mathcal{G}$; that is, $\mathcal{S}=\{S_1,S_2,\dots,S_m\}$ such that $S_i\subseteq \mathcal{G}$ for all $1\leq i\leq m$. The objective is to find a minimum-cardinality subset of $\mathcal{S}$ whose union is $\mathcal{G}$. Set Cover is known to be \np-hard~\cite{CormenLRS01} and even hard to approximate~\cite{Feige98}.

In this paper, we consider a geometric variant of the set cover problem that was first studied by Korman et al.~\cite{KormanPR18}. A set of line segments in the plane is said to be \emph{non-overlapping} if any two line segments from the set intersect in at most one point. Given a set $S$ of $n$ non-overlapping line segments in the plane, a \emph{cell} in the arrangement of $S$ is a maximally connected region that is not intersected by any line segment in $S$~\cite{KormanPR18}. Then, the objective of the \emph{Line Segment Covering} ($\mathsf{LSC}$) problem is to select a minimum number of line segments such that every cell in the arrangement of the line segments is covered. Here, a cell is \emph{covered} by a line segment if it is incident to the line segment (i.e., the line segment is in the set of line segments defining the boundary of the cell). We assume that at most two line segments may share a fixed point in the plane.

\paragraph{Related work.} Korman et al.~\cite{KormanPR18} proved that when the line segments are only horizontal and vertical, the \mfc problem is \np-hard and it remains \np-hard when the goal is to cover the ``rectangular'' cells. By a closer look at their hardness proof, one can see that the problem is \np-hard even if we are only allowed to select the line segments from one orientation (they only select vertical line segments when constructing a solution from a given truth assignment for the corresponding 3SAT problem). Moreover, the authors gave an $O(n\log n)$-time \fpt algorithm for covering the rectangular cells when parameterized by $k$, the size of an optimal solution. However, the algorithm does not work when the goal is to cover \emph{all} cells of the arrangement. The authors leave open studying the approximability of the problem.

The \mfc problem is closely related to a guarding problem studied by Bose et al.~\cite{BoseCCHKLT13}. Given a set of lines in the plane, they studied the problems of guarding cells of the arrangement by selecting a minimum number of lines, or guarding the lines by selecting a minimum number of cells. Here, ``guarding'' has the same meaning as ``covering'' in the \mfc problem. However, their results do not extend to the \mfc problem, because (as also noted by Korman et al.~\cite{KormanPR18}) they use some properties of lines that are not true for the case of line segments.

\paragraph{Our results.} In this paper, we prove the following results.
\begin{itemize}
\item We give a \ptas for the \mfc problem when the line segments in $S$ can have any arbitrarily orientations, but we are allowed to select the covering line segments from only one orientation. Given the \np-hardness of the problem~\cite{KormanPR18}, this settles the complexity of this variant of the problem.
\item When we allow selecting the covering line segments from more than one direction, we show that the \mfc problem is \apx-hard when the line segments in $S$ have two orientations and the goal is to cover the rectangular cells.
\item We give an \fpt algorithm for the \mfc problem when the line segments in $S$ have only two orientations and the goal is to cover \emph{all} cells of the arrangement. This complements the \fpt algorithm of Korman et al.~\cite{KormanPR18} as we do not restrict the covering only to rectangular cells.
\end{itemize}

\paragraph{Organization.} 
In Section \ref{sec:prelims}, we give some definitions and revisit some necessary background. We show our \ptas in Section~\ref{sec:ptas} and the \apx-hardness result in Section~\ref{sec:apxHard}. Finally, the \fpt algorithm is given in Section~\ref{sec:fpt} and we conclude the paper in Section~\ref{sec:conclusion}.

%%%%%%%%%%%% NEW SECTION %%%%%%%%%%%%%%%%%
\section{Preliminaries}
\label{sec:prelims}
In the following, we revisit some techniques and background that are used throughout this paper.

\paragraph{Local search.} Our \ptas for the \mfc problem is based on the local search technique, which was introduced independently by Mustafa and Ray~\cite{MustafaR10}, and Chan and Har-Peled~\cite{ChanH12}. Consider an optimization problem in which the objective is to compute a feasible subset $S'$ of a ground set $S$ whose cardinality is minimum over all such feasible subsets of $S$. Moreover, it is assumed that computing some initial feasible solution and determining whether a subset $S'\subseteq S$ is a feasible solution can be done in polynomial time. The local search algorithm for a minimization problem is as follows. Fix some fixed parameter $k$, and let $A$ be some initial feasible solution for the problem. In each iteration, if there are $A'\subseteq A$ and $M\subseteq S\setminus A$ such that $|A'|\leq k$, $|M|<|A'|$ and $(A\setminus A')\cup M$ is a feasible solution, then set $A=(A\setminus A')\cup M$ and re-iterate. The algorithm returns $A$ and terminates when no such local improvement is possible.

Clearly, the local search algorithm runs in polynomial time. Let $\mathcal{B}$ and $\mathcal{R}$ be the solutions returned by the algorithm and an optimal solution, respectively. The following result establishes the connection between local search technique and obtaining a $\mathsf{PTAS}$.
\begin{theorem}[\cite{ChanH12,MustafaR10}]
\label{thm:LSgivesPTAS}
Consider the solutions $\mathcal{B}$ and $\mathcal{R}$ for a minimization problem, and suppose that there exists a planar bipartite graph $H=(\mathcal{B}\cup\mathcal{R}, E)$ that satisfies the local exchange property: for any subset $\mathcal{B}'\subseteq\mathcal{B}$, $(\mathcal{B}\setminus\mathcal{B'})\cup N_H(\mathcal{B'})$ is a feasible solution, where $N_H(\mathcal{B'})$ denotes the set of neighbours of $\mathcal{B'}$ in $H$. Then, the local search algorithm yields a \ptas for the problem.
\end{theorem}

The local search was used by Mustafa and Ray~\cite{MustafaR10} to obtain a \ptas for geometric hitting set problem and by Chan and Har-Peled~\cite{ChanH12} to obtain a \ptas for geometric independent set problem. Since then, the technique has been used to get a \ptas for several other geometric problems, such as geometric dominating set~\cite{BandyapadhyayMMS18} and unique covering~\cite{AshokKMS15}.

\paragraph{Fixed-parameter tractability.} The theory of parameterized complexity was developed by Downey and Fellows~\cite{DowneyF99}. Let $\Sigma$ be a finite alphabet. Then, a \emph{parameterized} problem is a language $L\subseteq\Sigma^*\times\Sigma^*$ in which the second component is called the \emph{parameter} of the problem. A parameterized problem $L$ is said to be \emph{fixed-parameter tractable} or $\mathsf{FPT}$, if the question ``$(x_1,x_2)\in L?$'' can be decided in time $f(|x_2|)\cdot |x_1|^{O(1)}$, where $f$ is an arbitrary function. We call an algorithm with such running time $f(|x_2|)\cdot |x_1|^{O(1)}$, an \fpt algorithm.

For the rest of this paper, we denote a set of $n$ line segments in the plane by $S$ (i.e., $|S|=n$) and the resulting arrangement by $\mathcal{A}(S)$.

%%%%%%%%%%%% NEW SECTION %%%%%%%%%%%%%%%%%
\section{\ptas}
\label{sec:ptas}
In this section, we show that the \mfc problem admits a \ptas when the line segments in $S$ are in any orientations, but we can select line segments from only one orientation to cover the cells. To this end, we run the local search algorithm with parameter $k=c/\epsilon^2$ for some $\epsilon>0$, where $c$ is a constant. Let $\mathcal{B}$ be the solution returned by the algorithm and let $\mathcal{R}$ be an optimal solution. We can assume that $\mathcal{B}\cap\mathcal{R}=\emptyset$. This is because if $\mathcal{B}\cap\mathcal{R}\neq\emptyset$, then we can consider the sets $\mathcal{B}'=\mathcal{B}\setminus(\mathcal{B}\cap\mathcal{R})$ and $\mathcal{R}'=\mathcal{R}\setminus(\mathcal{B}\cap\mathcal{R})$, and analyze the algorithm with $\mathcal{B}'$ and $\mathcal{R}'$. Here, we mark the faces covered by a line segment in $\mathcal{B}\cap\mathcal{R}$ as ``covered'' so as we do not need to cover then in the new variant of the problem. This guarantees that the approximation factor of the original instance is upper bounded by that of the new instance with these two new sets $\mathcal{B}'$ and $\mathcal{R}'$.

\begin{figure}[t]
\centering
\includegraphics[width=0.80\textwidth]{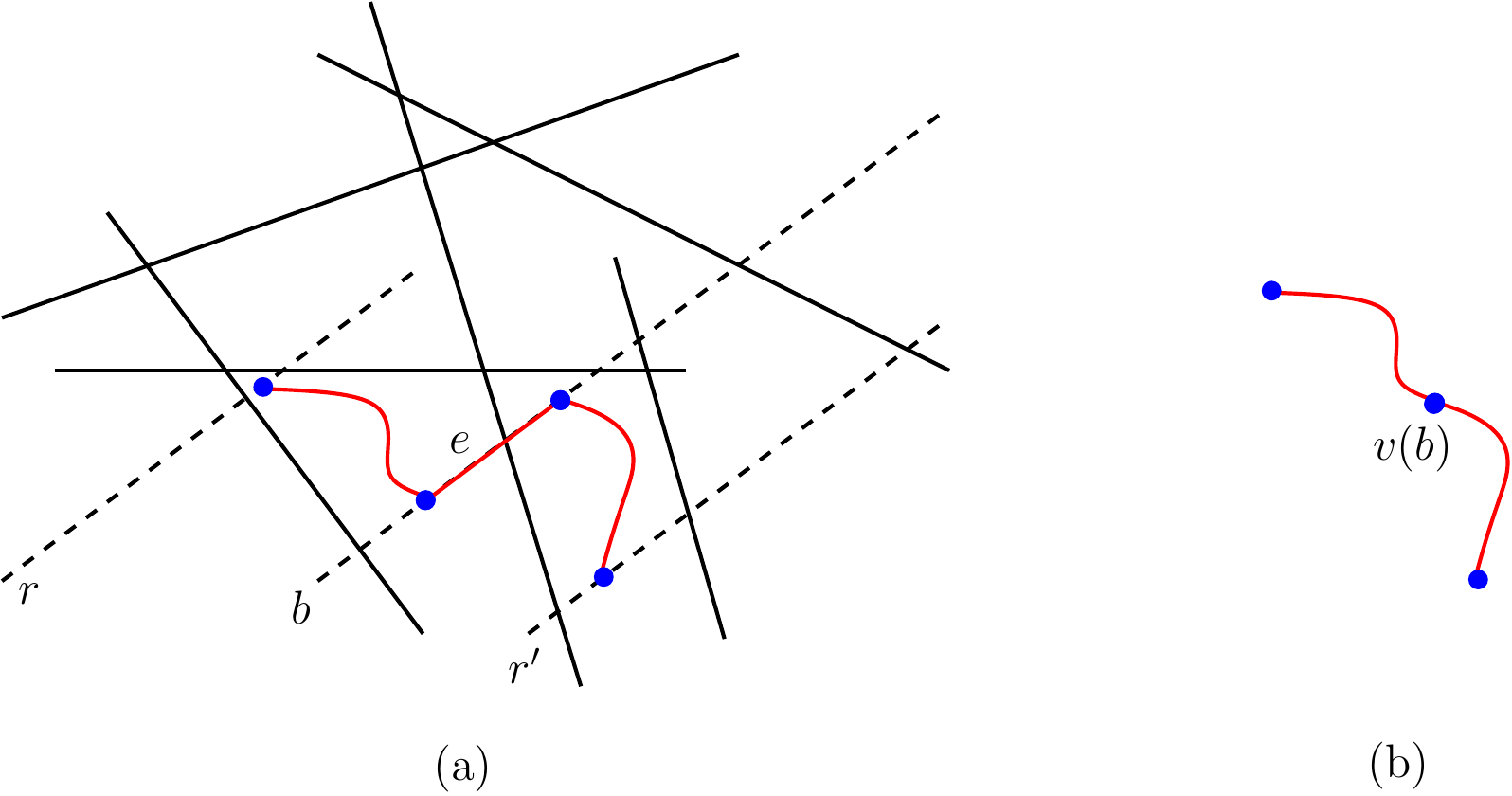}
\caption{(a) An example of three edges of graph $H'$. Here, $b\in\mathcal{B}$ and $r,r'\in\mathcal{R}$. (b) Two edges of graph $H$ by contracting the edge $e$ and obtaining one vertex $v(b)$ corresponding to line segment $b$.}
\label{fig:graphH}
\end{figure}

We now construct a planar bipartite graph $H=(\mathcal{B}\cup \mathcal{R}, E)$ that satisfies the local exchange property, hence proving that the problem admits a \ptas by Theorem~\ref{thm:LSgivesPTAS}. To this end, we first construct an auxiliary planar graph $H'$ and then show how to obtain $H$ from $H'$ by edge contraction. For each cell $f\in\mathcal{A}(S)$, let $b\in \mathcal{B}$ and $r\in \mathcal{R}$ be two line segments that cover $f$; we select a point $p\in b\cap f$ and $q\in r\cap f$ and connect them by a curve $c$ that lies in the interior of $f$ (except its endpoints $p$ and $q$). Notice that since both $\mathcal{B}$ and $\mathcal{R}$ are feasible solutions, we know that $\mathcal{B}$ contains at least one line segment that covers $f$ and $\mathcal{R}$ also contains at least one line segment that covers $f$, for all $f\in\mathcal{A}(S)$. We add $p$ and $q$ to $V(H')$ and $c$ to $E(H')$. We complete the definition of $H'$ by connecting every pair of consecutive points in $s\cap V(H')$, for all $s\in S$, by an edge that is exactly the portion of $s$ that lies between the pair of points. See Figure~\ref{fig:graphH}(a) for an example. Clearly, $H'$ is planar because the first set of edges are drawn in the interior of cells and each cell contains at most one edge. Moreover, the second set of edges are aligned with the line segments in $\mathcal{B}\cup \mathcal{R}$. Since the line segments in $\mathcal{B}\cup \mathcal{R}$ are non-overlapping and all have the same orientation, the second set of edges are also non-crossing. To obtain the graph $H$, for each segment $s\in \mathcal{B}\cup \mathcal{R}$, we contract the edges of $H'$ that are contained in $s$ such that we get a single point $v(s)$ corresponding to $s$; see Figure~\ref{fig:graphH}(b). So, $V(H)=\{v(s)|s\in \mathcal{B}\cup \mathcal{R}\}$. Graph $H$ is planar since $H'$ remains planar after this edge contraction. Moreover, $H$ is a bipartite graph as the edges of $H'$ with both endpoints belonging to a line segment in $\mathcal{B}$ or both endpoints belonging to a line segment in $\mathcal{R}$ are collapsed into a single point (i.e., $v(s)$).
\begin{lemma}
\label{lem:hIsPlanar}
Graph $H$ is planar and bipartite.
\end{lemma}

We next show that $H$ satisfies the exchange property.
\begin{lemma}
\label{lem:exchangeProperty}
Graph $H$ satisfies the local exchange property.
\end{lemma}
\begin{proof}
It is sufficient to show that for every cell $f\in\mathcal{A}(S)$, there are vertices $b\in \mathcal{B}$ and $r\in \mathcal{R}$ such that both segments corresponding to these vertices cover $f$ and $(b,r)\in E(H)$. Take any cell $f\in\mathcal{A}(S)$ and let $M\subseteq \mathcal{B}\cup \mathcal{R}$ be the set of all line segments that cover $f$. Notice that $M\cap \mathcal{B}\neq\emptyset$ and $M\cap \mathcal{R}\neq\emptyset$ because $\mathcal{B}$ and $\mathcal{R}$ are each a feasible solution. Then, by definition, there must be a $b\in M\cap \mathcal{B}$ and $r\in M\cap \mathcal{R}$ for which $(b,r)\in E(H)$. This completes the proof of the lemma.
\end{proof}

Putting everything together, we have the main result of this section.
\begin{theorem}
\label{thm:ptas}
There exists a \ptas for the line segment covering $(\mathsf{LSC})$ problem when the line segments in $S$ can have any orientation and we are allowed to select the covering line segments from only one orientation.
\end{theorem}

%%%%%%%%%%%% NEW SECTION %%%%%%%%%%%%%%%%%
\section{\apx-Hardness}
\label{sec:apxHard}
In this section, we show that the \mfc problem is \apx-hard when the line segments in $S$ have only two orientations and the goal is to cover the rectangular cells. To this end, we give an \elf-reduction from the Minimum Vertex Cover ($\mathsf{MVC}$) problem on graphs with maximum-degree three to this variant of the \mfc problem. Our reduction is inspired by the construction of Mehrabi~\cite{Mehrabi17}. As a reminder, we first give a formal definition of \elf-reduction~\cite{PapadimitriouY91}, which is one of the gap-preserving reductions. Let $\Pi$ and $\Pi'$ be two optimization problems with the cost functions $c_\Pi(.)$ and $c_{\Pi'}(.)$, respectively. We say that $\Pi$ \emph{\elf-reduces} to $\Pi'$ if there are two polynomial-time computable functions $f$ and $g$ such that the followings hold.
\begin{enumerate}
\item For any instance $x$ of $\Pi$, $f(x)$ is an instance of $\Pi'$.
\item If $y$ is a solution to $f(x)$, then $g(y)$ is a solution to $x$.
\item There exists a constant $\alpha>0$ such that
\[
OPT_{\Pi'}(f(x))\leq \alpha OPT_\Pi(x),
\]
where $OPT_Y(x)$ denotes the cost of an optimal solution for problem $Y$ on its instance $x$.
\item There exists a constant $\beta>0$ such that for every solution $y$ for $f(x)$,
\[
|OPT_\Pi(x)-c_\Pi(g(y))|\leq \beta |OPT_{\Pi'}(f(x))-c_\Pi(y)|,
\]
where $|x|$ denotes the absolute value of $x$.
\end{enumerate}

\begin{lemma}
\label{lem:mfcApx}
The minimum vertex cover $(\mathsf{MVC})$ problem on graphs with maximum-degree three is \elf-reducible to the \mfc problem, where $S$ is a set of horizontal and vertical line segments and the goal is to cover the rectangular cells of $\mathcal{A}(S)$.
\end{lemma}
\begin{proof}
\label{prf:lem:mfcApx}
Let $I$ be an instance of \mvc on graphs of maximum-degree three; let $G=(V, E)$ be the graph corresponding to $I$ and let $k$ be the size of the smallest vertex cover in $G$. First, let $u_1, \dots, u_n$ be an arbitrary ordering of the vertices of $G$, where $n=\lvert V\rvert$. In the following, we give a polynomial-time computable function $f$ that takes $I$ as input and outputs an instance $f(I)$ of the \mfc problem.

We first describe the vertex gadgets. For each vertex $u_i$, $1\leq i\leq n$, construct a horizontal line segment $H_i$ and a vertical line segment $V_i$, and connect them as shown in Figure~\ref{fig:mfcApx}. We call the (blue) horizontal line segment used in the connection of $H_i$ and $V_i$ the \emph{horizontal connector} $C_i$ of $i$. Moreover, there are four (small, dashed) line segments used in the connection of $H_i$ and $V_i$ that we call the \emph{small connectors} of $i$. Notice that these five ``connectors'' along with $H_i$ and $V_i$ form exactly two rectangular cells. For each edge $(u_i, u_j)\in E$, where $i<j$, we add two small line segments, one horizontal and one vertical, at the intersection point of $V_i$ and $H_j$ such that they intersect each other as well as each intersects one of $V_i$ and $H_j$, hence forming a rectangular cell; see the two (red, dashed) line segments at the intersection of $V_1$ and $H_2$ in Figure~\ref{fig:mfcApx} for an example. We call such a pair \emph{edge line segments} and denote them by $E_{i,j}$. Finally, for every rectangular cell whose four sides are \emph{all} defined by the line segments corresponding to a 4-subset of $\{H_i,V_i| 1\leq i\leq n\}$ (i.e., the cell is not covered by a horizontal connector or edge line segments), we insert a vertical line segment into the cell so as to make it non-rectangular; see the vertical (red) line segment in Figure~\ref{fig:mfcApx}. This ensures that every rectangular cell is incident either to a horizontal connector or to edge line segments $E_{i,j}$ for some $i$ and $j$. This gives the instance $f(I)$ of the \mfc problem. Notice that $f$ is a polynomial-time computable function. In the following, we denote an optimal solution for the instance $X$ of a problem by $s^*(X)$. We now prove that all the four conditions of \elf-reduction hold.

First, let $M$ be a vertex cover of $G$ of size $k$. Denote by $H[M] = \{H_i | u_i\in M\}$ the set of horizontal line segments induced by $M$ and define $V[M]$ analogously. Moreover, let $C[M] = \{C_i | u_i \notin M\}$ be the set of horizontal connectors whose corresponding vertex is not in $M$. We show that $F = H[M]\cup V[M]\cup C[M]$ is a feasible solution for covering all the rectangular cells of $f(I)$. Let $f$ be a rectangular cell. Then, $f$ must be incident either to a horizontal connector or to edge line segments $E_{i,j}$ for some $i$ and $j$. First, if $f$ is incident to a horizontal connector $C_i$, then either $C_i\in F$ or $H_i\in F$ and $V_i\in F$ by the construction of $F$ and so $f$ is covered either way. Next, if $f$ is incident to edge line segments $E_{i,j}$ for some $i$ and $j$, where w.l.o.g. $i<j$, then either $V_i\in F$ or $H_j\in F$ because we know that $u_i\in M$ or $u_j\in M$. So, $f$ is again covered in this case. Therefore, $F$ is a feasible solution.

\begin{figure}[t]
\centering
\includegraphics[width=0.80\textwidth]{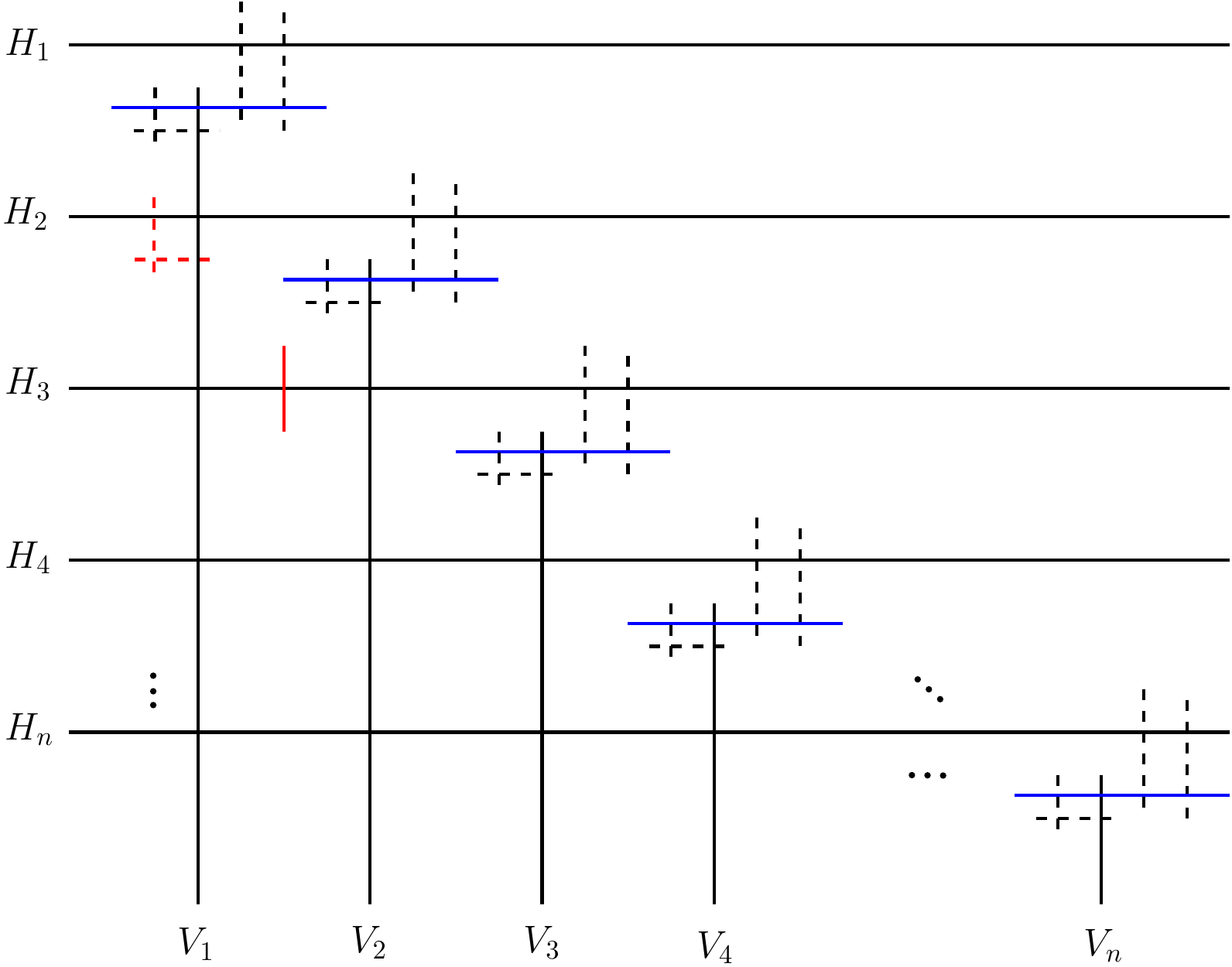}
\caption{An illustration in support of the construction in the proof of Lemma~\ref{lem:mfcApx}.}
\label{fig:mfcApx}
\end{figure}

Second, let $F$ be any feasible solution for $f(I)$. Notice that we can construct a feasible solution $F'$ for $f(I)$ such that $\lvert F'\rvert\leq\lvert F\rvert$ and $F'$ consists of only $H_i$ and $V_i$ for some $i$, or a horizontal connector. This is because \begin{inparaenum}[(i)] \item any rectangular cell covered by a small connector is also covered by a horizontal connector, and \item\label{inl:domiantingEij} any cell covered by a pair of edge line segments $E_{i,j}$ (for some $i$ and $j$) is also covered by $V_i$ and $H_j$. \end{inparaenum} For~\eqref{inl:domiantingEij}, if exactly one of the line segments in $E_{i,j}$ is in $F$, then we replace it with exactly one of $V_i$ or $H_j$. Otherwise, if both line segments of $E_{i,j}$ are in $F$, then we replace both of them with $V_i$ and $H_j$. So, $\lvert F'\rvert\leq\lvert F\rvert$ and $F'$ is a feasible solution for $f(I)$. Now, let $M = \{u_i | H_i\in F' \mbox{ or } V_i\in F'\}$. To show that $M$ is a vertex cover for $G$, consider any edge $(u_i,u_j)\in E$, where $i<j$. Then, we know that there exists a rectangular cell at the intersection of $V_i$ and $H_j$ that must be covered by $F'$. Since none of the two edge line segments of $E_{i,j}$ are in $F'$, we conclude that at least one of $V_i$ and $H_j$ is in $F'$, which means that $u_i\in M$ or $u_j\in M$. Hence, $M$ is a vertex cover.

Third, observe that $\lvert H[M]\rvert=\lvert V[M]\rvert=\lvert M\rvert=k$ and also $\lvert C[M]\rvert=n-k$. Given that $G$ has degree three, $k\geq n/4$ and so $\lvert s^*(f(I))\rvert\leq n-k+k+k\leq 5k\leq 5\lvert s^*(I)\rvert$.

We now prove the last condition of \elf-reduction. First, define $\both[F']=\{H_i,V_i | H_i,V_i \in F'\}$; that is, the paths of a vertex $u_i$, where \emph{both} its horizontal and vertical line segments appear in $F'$. Also, define $\one[F']$ to be the remaining line segments corresponding to either $H_i$ or $V_i$ for some $i$; i.e., those of $u_i$, where \emph{only} one of its line segments appears in $F'$. Take any vertex $i$. To cover the two rectangular cells incident to the horizontal connector of $i$, we must have $C_i\in F'$ or $H_i, V_i\in F'$; this is true for all $i$. Then, $\lvert C[F']\rvert+\lvert \both[F']\rvert/2\geq n$. Moreover, $\lvert \one[F']\rvert+\lvert \both[F']\rvert/2\geq k$ since $M$ is a vertex cover of $G$. Therefore, $\lvert F'\rvert\geq\lvert \both[F']\rvert+\lvert \one[F']\rvert+\lvert C[F']\rvert\geq\lvert \one[F']\rvert+\lvert \both[F']\rvert/2+n\geq k+n$. By this and our earlier inequality $\lvert s^*(f(I))\rvert\leq n-k+k+k$, we have $\lvert s^*(f(I))\rvert=n+k$. Now, suppose that $\lvert F\rvert=\lvert s^*(f(I))\rvert+c$ for some $c\geq 0$. Then,
\begin{align*}
& \lvert F\rvert-\lvert s^*(f(I))\rvert=c\\
& \Rightarrow \lvert F\rvert-(n+k)=c\\
& \Rightarrow \lvert F'\rvert-(n+k)\leq c\\
& \Rightarrow \lvert \one[F']\rvert+\lvert \both[F']\rvert/2+n-(n+k)\leq c\\
& \Rightarrow \lvert \one[F']\rvert+\lvert \both[F']\rvert/2-k\leq c\\
& \Rightarrow \lvert M\rvert-\lvert s^*(I)\rvert\leq c.
\end{align*}
That is, $\lvert M\rvert-\lvert s^*(I)\rvert\leq\lvert F\rvert-\lvert s^*(f(I))\rvert$. This concludes our \elf-reduction from \mvc on graphs of maximum-degree three to \mfc with $\alpha=5$ and $\beta=1$.
\end{proof}

\begin{theorem}
\label{thm:mfcApx}
The line segment covering $(\mathsf{LSC})$ problem is \apx-hard when the line segments in $S$ are either horizontal or vertical and the goal is to cover the rectangular cells of $\mathcal{A}(S)$.
\end{theorem}

%%%%%%%%%%%% NEW SECTION %%%%%%%%%%%%%%%%%
\section{\fpt}
\label{sec:fpt}
In this section, we show that the \mfc problem is fixed-parameter tractable (parametrized by the size of an optimal solution) when the line segments in $S$ are either horizontal or vertical, and the goal is to cover all the cells in $\mathcal{A}(S)$. This complements the \fpt result of Korman et al.~\cite{KormanPR18}, where the goal is to cover the rectangular cells. Throughout this section, let $k$ be the size of an optimal solution.

Our \fpt follows the framework of Korman et al.~\cite{KormanPR18}. That is, we formulate the \mfc problem as a hitting set problem and argue that we only need to hit an $O(k^3)$ number of sets; hence, obtaining a kernel of size $O(k^3)$ for the problem. The \fpt of Korman et al.~\cite{KormanPR18} is based on the fact that any three orthogonal line segments can cover at most two ``rectangular'' cells (i.e., at most two rectangular cells can be incident to all the three line segments). As an analogous result, we prove in Lemma~\ref{lem:3segments} that the number of such cells can be at most six when the goal is to cover \emph{all} cells, including non-rectangular ones. We will then apply this result to obtain the desired kernel.
\begin{lemma}
\label{lem:3segments}
Let $S$ be a set of $n$ axis-parallel line segments in the plane. Then, for any three line segments $s_1,s_2,s_3\in S$, there are at most six cells in $\mathcal{A}(S)$ that can be covered by all three line segments $s_1,s_2$ and $s_3$.
\end{lemma}

\begin{figure}[t]
\centering
\includegraphics[width=0.95\textwidth]{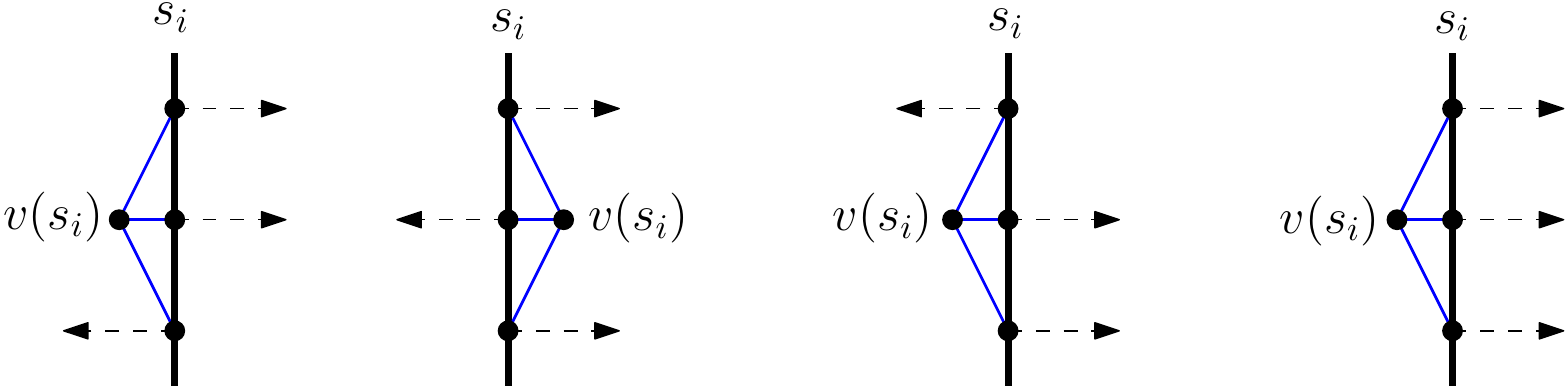}
\caption{Placing a new vertex $v(s_i)$ close to $s_i$ and connecting it to the vertices corresponding to the incident cells. The arrows indicate the side on which the cell lies.}
\label{fig:disjointK33}
\end{figure}

\begin{proof}
Take any three line segments $s_1,s_2$ and $s_3$ in $\mathcal{A}(S)$ and let $F$ be the set of all cells in $\mathcal{A}(S)$ that are covered by all three line segments $s_1,s_2$ and $s_3$. We need to show that $|F|\leq 6$. To this end, we construct a planar graph $H$ corresponding to $s_1,s_2, s_3$ and the cells in $F$ and will then argue that this graph must contain a subdivision of $K_{3,3}$ if $|F|>6$. We next give the details. Let $f$ be a cell in $F$. Consider a point $p(f)$ in the interior of $f$ as well as a distinct point $p(s_i,f)$ in $s_i\cap f$, for all $i=1,2,3$ (notice that $p(s_i,f)$ is on the boundary of $f$). These points together form the set of vertices of $H$; that is, $V(H)=\{p(f):f\in F\}\cup\{p(s_i,f):i=1,2,3, f\in F\}$. Now, for each $i=1,2,3$, consider an ordering of the points $p(s_i, f)$ on $s_i$, $f\in F$, and connect every two consecutive points by an edge, which is exactly the portion of $s_i$ that lies between the two points. Moreover, for each cell $f$, we connect $p(f)$ to $p(s_i,f)$ by a curve that lies strictly in the interior of $f$ (except at its endpoints) for all $i=1,2,3$. Then, the edge set $E(H)$ of $H$ consists of the set of all edges connecting the consecutive points as we as the curves $(p(f),p(s_i,f))$, for all $f\in F$ and $i=1,2,3$. Clearly, $H$ is a planar graph. In the following, we consider several cases depending on whether the line segments $s_1,s_2$ and $s_3$ intersect each other; observe that there can be at most two intersection points between them.

\paragraph{Case 1.} There is no intersection point; that is, the line segments $s_1, s_2$ and $s_3$ are pairwise disjoint. In this case, we show that in fact $|F|\leq 2$. To this end, suppose for a contradiction that $|F|>2$. Take any three cells $f_1,f_2,f_3\in F$ and consider the subgraph $H'$ of $H$ induced by $\{p(f_i):i=1,2,3\}\cup\{p(s_i,f_j):i,j=1,2,3\}$. Now, consider the graph $G$ constructed from $H'$ as follows. For each $s_i$, $i=1,2,3$, we place a new vertex $v(s_i)$ close to $s_i$ and connect it to the three vertices $p(s_i,f_j)$ for all $j=1,2,3$ such that the resulting graph remains planar. One can easily verify that this is doable since the three line segments are disjoint and so there are a few cases for where to place $v(s_i)$ depending on which side of $s_i$ the three cells lie; see Figure~\ref{fig:disjointK33}. Observe that the resulting graph $G$ is a planar drawing of a subdivision of $K_{3,3}$, which is not possible. So, $|F|\leq 2$.

\paragraph{Case 2.} There is exactly one intersection point; assume w.o.l.g. that $s_1$ is horizontal, $s_2$ and $s_3$ are vertical and $s_1$ intersects $s_2$. Here, we show that $|F|\leq 4$. Again, suppose for a contradiction that $|F|>4$. Then, considering the graph $H$, there must be at least three vertices in $\{p(s_1,f):f\in F\}$ that lie w.l.o.g. to the right of $s_2$. Take any three such vertices and denote the corresponding cells by $f_1,f_2,f_3$. We can now construct the graph $G$ analogous to the one in Case 1 with these three cells and so obtain a planar drawing of a subdivision of $K_{3,3}$, which is a contradiction.

\paragraph{Case 3.} There are two intersection points. Here, we show that $|F|\leq 6$ and we use a similar argument to those in the previous cases. Denote the endpoints of $s_1$ by $a$ and $b$, and let $p$ and $q$ be the intersection points of $s_1$ with $s_2$ and $s_3$; assume w.l.o.g. that $a$ is the left endpoint of $s_1$ and that $p$ lies to the left of $q$. If $|F|>7$, then at least one of the line segments $ap, pq$ and $qb$ must contain three vertices of $\{p(s_1,f):f\in F\}$; assume w.l.o.g. that it is $pq$. Then, take any three such vertices on $pq$ and consider the three cells $f_1,f_2$ and $f_3$ corresponding to these vertices. We can now construct the graph $G$ analogous to the one in Case 1 and so obtain a planar drawing of a subdivision of $K_{3,3}$, which is a contradiction. As such, $|F|\leq 6$.

By the three cases described above, we conclude that $|F|\leq 6$.
\end{proof}

\begin{figure}[t]
\centering
\includegraphics[width=0.45\textwidth]{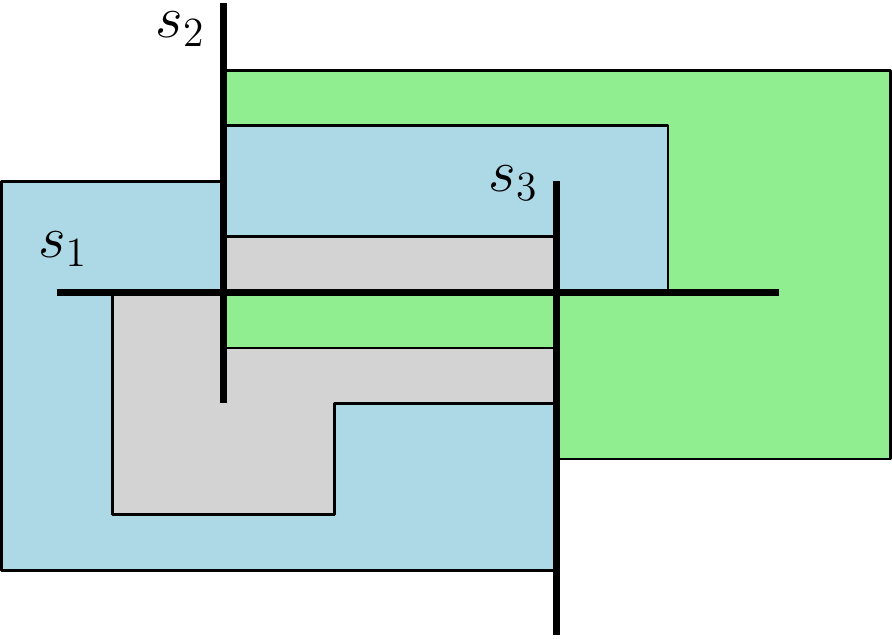}
\caption{Three line segments $s_1$, $s_2$ and $s_3$ cover six cells.}
\label{fig:3segments}
\end{figure}

We note that the upper bound in Lemma~\ref{lem:3segments} is tight as Figure~\ref{fig:3segments} shows an example with three line segments that cover six cell. We now apply Lemma~\ref{lem:3segments} to obtain our $\mathsf{FPT}$. We first formulate the \mfc problem as a hitting set problem as follows. The ground set is $S$ and for each cell in $\mathcal{A}(S)$, there exists a set that contains the line segments that cover the cell. Let $\mathcal{C}$ be the resulting set of subsets of $S$. Then, the \mfc problem is equivalent to selecting a minimum number of elements from $S$ such that each set in $\mathcal{C}$ is hit by at least one selected element.

We first reduce the set $\mathcal{C}$ to a set $\mathcal{C}_1$ as follows. For every pair of line segments $\{s_i,s_j\}\in S$, if they appear in more than $6k$ sets $\mathcal{C}$, then we remove all such sets form $\mathcal{C}$ and add the set $\{s_i,s_j\}$ to $\mathcal{C}$. Let $\mathcal{C}_1$ be the resulting set.
\begin{lemma}
\label{lem:firstReduction}
A set $S'\subseteq S$ with $|S'|\leq k$ is a minimum-size cover of $\mathcal{C}$ if and only if it is a minimum-size cover of $\mathcal{C}_1$.
\end{lemma}
\begin{proof}
We prove the lemma by an argument similar to the one by Korman et al.~\cite{KormanPR18}. The lemma clearly follows if $\mathcal{C}=\mathcal{C}_1$. So, assume that $X=\mathcal{C}_1\setminus\mathcal{C}$ and $Y=\mathcal{C}\setminus\mathcal{C}_1$ are two non-empty sets. Let $S'$ with $|S'|\leq k$ be a minimum-size cover for $\mathcal{C}_1$. First, $S'$ is also a cover for $\mathcal{C}$ because for every set $M\in Y$ there exists a pair of line segment $s_i$ and $s_j$ such that both $s_i$ and $s_j$ are in $M$ and we have $\{s_i,s_j\}\in X$. We now prove that $S'$ is also a cover of minimum size for $\mathcal{C}$.

Suppose for a contradiction that there exists a cover $S''$ for $\mathcal{C}$ such that $|S''|<|S'|$. Then, $S''$ cannot be a cover $\mathcal{C}_1$ because $S'$ is a cover of minimum size for $\mathcal{C}_1$. Since $S''$ covers $\mathcal{C}\cap\mathcal{C}_1$, there must exist $\{s_i,s_j\}\in X$ such that neither $s_i$ nor $s_j$ is in $S''$. But, we introduced the set $\{s_i,s_j\}$ into $\mathcal{C}_1$ because there were more than $6k$ sets containing both $s_i$ and $s_j$. If neither $s_i$ nor $s_j$ is in $S''$, then every other line segment can cover at most six of such subsets by Lemma~\ref{lem:3segments}. Therefore, $|S''|>k$ --- a contradiction. By a similar argument, we can show that a minimum-size cover of $\mathcal{C}$ is also a minimum-size cover for $\mathcal{C}_1$. This completes the proof of the lemma.
\end{proof}

Next, we reduce $\mathcal{C}_1$ to a new set $\mathcal{C}_2$ as follows. For each line segment $s\in S$, we count how many sets in $\mathcal{C}_1$ contain $s$. If $s$ appears in more than $6k^2$, then we remove all those sets and add the set $\{s\}$ to $\mathcal{C}_1$. Let $\mathcal{C}_2$ denote the resulting set.
\begin{lemma}
\label{lem:secondReduction}
A set $S'\subseteq S$ with $|S'|\leq k$ is a minimum-size cover for $\mathcal{C}_1$ if and only if it is a minimum-size cover for $\mathcal{C}_2$.
\end{lemma}
\begin{proof}
The lemma follows if $\mathcal{C}_1=\mathcal{C}_2$. So, assume that $X'=\mathcal{C}_2\setminus\mathcal{C}_1$ and $Y'=\mathcal{C}_1\setminus\mathcal{C}_2$ are two non-empty sets. Let $S'$ with $|S'|\leq k$ be a minimum-size cover for $\mathcal{C}_2$. For any set $M\in Y'$, there exists a singleton set in $X'$ whose member is in $M$. This means that $S'$ is also a cover for $\mathcal{C}_1$. We next show that $S'$ is also a minimum-size cover for $\mathcal{C}_1$.

Suppose for a contradiction that there exists a cover $S''$ for $\mathcal{C}_1$ such that $|S''|<|S'|$. Therefore, $S''$ is not a cover of $\mathcal{C}_2$. Since $S''$ covers $\mathcal{C}_1\cap\mathcal{C}_2$, there must exist a set in $X'$ that is not covered by $S''$. Notice that this set must be of size 1 from the construction of $\mathcal{C}_2$; let $\{s\}$ be such a set, where $s\in S$. The reason we have the set $\{s\}$ in $\mathcal{C}_2$ is that because there were more than $6k^2$ sets in $\mathcal{C}_1$ containing $s$. If $s$ is not in $S''$, then all such sets of $\mathcal{S}_1$ must be covered by other line segments. But, from the construction of $\mathcal{C}_1$, every pair of line segments can appear in at most $6k$ sets. So, $|S''|$ must be greater than $k$, which is a contradiction. A similar argument can be used to show that a minimum-size cover for $\mathcal{C}_1$ is also is minimum-size cover for $\mathcal{C}_2$. This completes the proof of the lemma.
\end{proof}

Consider the set $\mathcal{C}_2$. By Lemma~\ref{lem:secondReduction}, no line segment of $S$ appears in more than $6k^2$ sets in $\mathcal{C}_2$. Therefore, if $|\mathcal{C}_2|>6k^3$, then the problem does not have a cover of size at most $k$. Since the construction of $\mathcal{C}_2$ can be done in polynomial time, we have the following result.
\begin{lemma}
\label{lem:kernel}
For the \mfc problem on a set of axis-parallel line segments, in polynomial time, we can either obtain a kernel of size $O(k^3)$ or conclude that the problem does not have a cover of size at most $k$, where $k$ is the size of an optimal cover.
\end{lemma}

Since having a kernel of size $O(k^3)$ implies that the problem is $\mathsf{FPT}$~\cite{DowneyF13}, we have the main result of this section.
\begin{theorem}
\label{thm:fpt}
The line segment covering $(\mathsf{LSC})$ problem on a set of axis-parallel line segments is \fpt with respect to the size of an optimal cover.
\end{theorem}

%%%%%%%%%%%% NEW SECTION %%%%%%%%%%%%%%%%%
\section{Conclusion}
\label{sec:conclusion}
In this paper, we considered the problem of covering the cells in the arrangement of a set of line segments in the plane. We proved that the problem admits a \ptas when the covering line segments can be selected from only one orientation. We then showed that if we allow selecting the covering line segments from more than one orientation, then the problem is \apx-hard when we are interested in covering the rectangular cells. Finally, we gave an \fpt algorithm for the problem when the line segments have only two orientations, but the goal is to cover all the cells. Our \apx-hardness rules out the possibility of a \ptas for ``covering rectangular faces'' variant of the problem, but is there a 2-approximation algorithm for the problem? For the more general variant, where the line segments are in any orientation, covering line segments can be selected from any orientation and the goal is to cover all the cells, can we obtain a $c$-approximation algorithm for some small constant $c$?

%%%%%%%%%%%% NEW SECTION %%%%%%%%%%%%%%%%%
\bibliographystyle{plain}
\bibliography{ref}

\end{document}